\documentclass[fleqn,10pt]{wlscirep}
\usepackage[T1]{fontenc}

\title{A Simple Multiscale Intermediate Coupled Stochastic Model for El Ni\~no Diversity and Complexity}

\author[1]{Nan Chen}
\author[2,3,4,*]{Xianghui Fang}

\affil[1]{Department of Mathematics, University of Wisconsin-Madison, Madison, WI, USA}
\affil[2]{Department of Atmospheric and Oceanic Sciences and Institute of Atmospheric Sciences, Fudan University, 220 Handan Rd., Shanghai 200433, China}
\affil[3]{Innovation Center of Ocean and Atmosphere System, Zhuhai Fudan Innovation Research Institute, Zhuhai 518057, China}
\affil[4]{CMA-FDU Joint Laboratory of Marine Meteorology, Shanghai 200438, China}

\affil[*]{fangxh@fudan.edu.cn}

\begin{abstract}
El Ni\~no-Southern Oscillation (ENSO) is the most prominent interannual climate variability in the tropics and exhibits diverse features in spatiotemporal patterns. In this paper, a simple multiscale intermediate coupled stochastic model is developed to capture the ENSO diversity and complexity. The model starts with a deterministic and linear coupled interannual atmosphere, ocean and sea surface temperature (SST) system. It can generate two distinct dominant linear solutions that represent the eastern Pacific (EP) and the central Pacific (CP) El Ni\~nos, respectively. In addition to adopting a stochastic model for characterizing the intraseasonal wind bursts, another simple stochastic process is developed to describe the decadal variation of the background Walker circulation. The latter links the two dominant modes in a simple nonlinear fashion and advances the modulation of the strength and occurrence frequency of the EP and the CP events. Finally, a cubic nonlinear damping is adopted to parameterize the relationship between subsurface temperature and thermocline depth. The model succeeds in reproducing the spatiotemporal dynamical evolution of different types of the ENSO events. It also accurately recovers the strongly non-Gaussian probability density function, the seasonal phase locking, the power spectrum and the temporal autocorrelation function of the SST anomalies in all the three Ni\~no regions (3, 3.4 and 4) across the equatorial Pacific. Furthermore, both the composites of the SST anomalies for various ENSO events and the strength-location bivariate distribution of equatorial Pacific SST maxima for the El Ni\~no events from the model simulation highly resemble those from the observations.
\end{abstract}
\begin{document}

\flushbottom
\maketitle
%
%
\thispagestyle{empty}


\section{Introduction}
El Ni\~no-Southern Oscillation (ENSO) is the most prominent interannual variability in the tropics. It also affects the global climate, ecosystem and socioeconomic development through the atmospheric teleconnections \cite{ropelewski1987global,mcphaden2006enso}. Therefore, understanding and predicting ENSO is a central problem in contemporary meteorology with large societal impacts. Bjerknes \cite{bjerknes1969atmospheric} first suggested that ENSO is the product of tropical air-sea interaction. Since then, great achievements have been made in its simulation and prediction abilities \cite{latif1998review, neelin1998enso}.

From the traditional point of view, El Ni\~no is defined as the anomalous warm sea surface temperature (SST) in the equatorial eastern Pacific (EP) region. Zebiak and Cane \cite{zebiak1987model} developed the first coupled ocean-atmosphere model of intermediate complexity that successfully characterizes and predicts these EP warming events. Several deterministic and linear conceptual models were also proposed to explain the slow physics of ENSO. Among them, the delayed oscillator \cite{schopf1988vacillations} describes the delayed effects of oceanic wave reflection at the ocean western boundary on the EP SST anomalies while the recharge-discharge oscillator \cite{jin1997equatorial} combines SST dynamics and ocean adjustment dynamics into a coupled basinwide recharge oscillator that relies on the non-equilibrium between the zonal mean equatorial thermocline depth and wind stress.

\begin{figure}
    \centering\hspace*{0cm}
    \includegraphics[width=12cm]{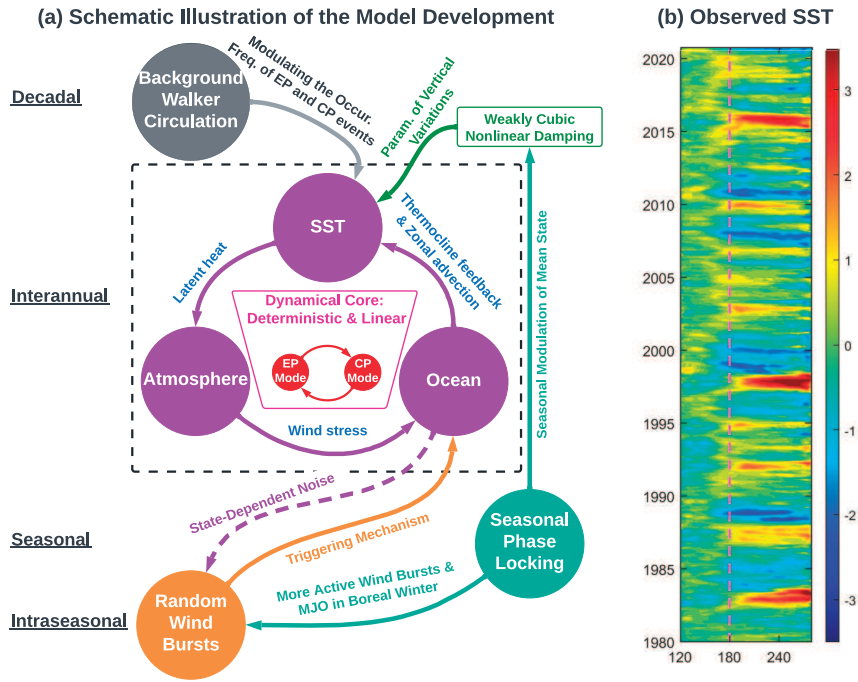}
    \caption{Panel (a): A schematic illustration of the multiscale model developed here. Panel (b): The observational SST anomaly from 1980 to 2020 (unit: $^o$C). It is based on the GODAS dataset \cite{behringer2004evaluation} and is computed by averaging over 5$^o$S to 5$^o$N followed by removing the monthly mean climatology of the entire period.}
    \label{Schematic}
\end{figure}

With the continuously improved understanding of nature, the spatiotemporal diversity and complexity of ENSO have been progressively highlighted \cite{capotondi2015understanding,timmermann2018nino}. In particular, the observational data shows that the center of anomalous SST is mainly located in the EP from 1980 to 2000, whereas it lies more towards the central Pacific (CP) after 2000 \cite{ashok2007nino, kao2009contrasting, kim2012statistical}. See Panel (b) of Figure \ref{Schematic}. The emergence of these distinct ENSO events is called the El Ni\~no diversity \cite{capotondi2015understanding}. It suggests the existence of at least two types of El Ni\~nos, which are named as the EP and the CP 
ones when the peak of the SST anomaly locates in the cold tongue and near the dateline region, respectively \cite{larkin2005global,yu2007decadal,ashok2007nino,kao2009contrasting,kug2009two}. It is important to notice that the shift of the warming center can make significant differences in the air-sea coupling over the equatorial Pacific, which changes the way ENSO affects the global climate and brings serious challenges to its prediction \cite{chen2008nino, jin2008current, barnston2012skill, hu2012analysis, zheng2014asymmetry, fang2015cloud, sohn2016strength, santoso2019dynamics}. In addition to these two major categories, individual ENSO events further exhibit diverse characteristics in spatial pattern, peak intensity, and temporal evolution, which are known as the ENSO complexity \cite{timmermann2018nino}. Thus, developing effective dynamical models that capture the ENSO complexity is of practical importance, not only for improving the understanding of the formation mechanisms of ENSO but advancing the prediction of different ENSO events and the associated varying climatic impacts as well.

The main physical mechanisms of the EP and the CP El Ni\~nos are very different.
Due to the strong zonal asymmetry of the tropical Pacific air-sea system, the thermocline, which is a thin layer that separates the upper warm water from the cold water in the lower layer, is deep in the western Pacific (WP) while shallow in the EP region. Such a structure is consistent with the easterly trade wind. As a result, the SST in the EP is more susceptible to the oceanic vertical processes, i.e., the thermocline feedback. On the other hand, the background mean state suggests that the CP is the region with the largest zonal SST gradient. Consequently, the anomalous zonal current can significantly affect the local SST variations, which means the development of the CP type of ENSO is largely influenced by the zonal advective feedback \cite{kug2009two,kug2010warm,ham2012well,kug2012improved,chen2016simple,fang2018three,fang2018simulating}. It is worthwhile to point out that, since the state-of-art coupled general circulation models (CGCMs) in general have difficulties in accurately describing the mean state of the tropical Pacific (e.g., the unrealistic westward extension of the cold tongue), the simulations often contain biases in reproducing the equatorial SST gradient and the relevant zonal advective feedback \cite{lin2007double,fang2018simulating,xie2018two,planton2021evaluating}. Such an issue is one of the main reasons that result in great challenges for the CGCMs to correctly simulate the El Ni\~no diversity \cite{wittenberg2009historical,ham2012well,capotondi2015understanding,fang2015cloud,fang2018simulating}.

In addition to the interannual state variables, it is essential to further take into consideration several other key variabilities belonging to different temporal scales in the modeling procedure to realistically generate the ENSO complexity. On one hand, the intraseasonal atmospheric variability, e.g., the westerly wind burst (WWB) \cite{harrison1997westerly, vecchi2000tropical, tziperman2007quantifying} and the Madden-Julian oscillation (MJO) \cite{hendon2007seasonal, puy2016modulation}, has been understood as one of the main sources that lead to the ENSO irregularity and extreme events. Specifically, WWBs can influence the ENSO development by stimulating eastward-propagating oceanic Kelvin waves, generating surface zonal currents and weakening the evaporation. In fact, many modeling works have been attempted to incorporate  semistochastic parameterization for the WWBs  \cite{gebbie2007modulation, tziperman2007quantifying, gebbie2009predictability, levine2016extreme, thual2016simple,thual2017seasonal} and suggest that the coupled feedbacks between the interannual SST and the intraseasonal WWBs is sufficient to transfer a damped system to a semi-regular self-sustained oscillator. Likewise, in light of an intermediate coupled model (ICM), Lian et al. \cite{lian2014theoretical} found that the WWBs are responsible for the existence of the irregularity and intensity of El Ni\~no, whose specific characteristics depend on the timing of the WWBs relative to the phase of the recharge–discharge cycle.
On the other hand, ENSO is also modulated by the decadal variation of the background mean state. Particularly, McPhaden et al. \cite{mcphaden2011nino} and Xiang et al. \cite{xiang2013new} revealed that the changes in the equatorial Pacific around 2000, i.e., a La Ni\~na-like background state with enhanced trade winds and a more tilted thermocline, are in favor of the occurrence of more frequent CP El Ni\~no events. Power et al. \cite{power2021decadal} also emphasized the role of the decadal variability in affecting the equatorial Pacific. In addition, by extending the recharge oscillator into a three-region (i.e., WP, CP and EP) conceptual model that contains a set of 6 stochastic ordinary differential equations and includes both the thermocline and zonal advective feedbacks, Chen et al. \cite{chen2022multiscale} demonstrated that the decadal variability plays a crucial role in modulating the occurrence of the CP and EP El Ni\~nos.

The conceptual model in Chen et al. \cite{chen2022multiscale} succeeds in capturing many desirable large-scale features of the ENSO complexity. It thus provides an essential theoretical basis for the development of a more sophisticated dynamical model, namely an ICM, that aims at realistically reproducing detailed spatiotemporal patterns of the ENSO complexity. Different from the conceptual models, the ICMs have unique advantages of incorporating more elaborate underlying physics and spatially-extended dynamics into the model development that facilitate the understanding and prediction of nature. The ICMs also serve as a bridge that connects the low-order conceptual models and the more complicated CGCMs with a relatively low computational cost.

To this end, a stochastic ICM for the ENSO complexity is developed in this paper. The dynamical core of this new stochastic ICM is deterministic and linear, which involves a coupled interannual atmosphere, ocean and SST system that drives the basic ENSO dynamics in a simple fashion. Here, the latent heating that is proportional to the SST  is depleted from the ocean and forces an atmospheric circulation. The resulting zonal wind stress in turn forces ocean dynamics that provides a feedback to the SST through the thermocline depth anomalies and the ocean zonal advection. The coupled linear and deterministic interannual starting model can generate two distinct dominant linear solutions that represent the EP and the CP El Ni\~nos, respectively, which are essential for simulating the ENSO complexity. The interannual components are then coupled with  the intraseasonal and the decadal variabilities, which are both described by suitable stochastic processes. The former is the main contributor to the ENSO irregularity and extreme events while the latter links the two dominant modes in a simple nonlinear fashion and advances the modulation of the strength and occurrence frequency of the EP and the CP events. Seasonal synchronization is further incorporated into the model, facilitating the ENSO events to have a tendency to peak in boreal winter. Finally, a cubic nonlinear damping is adopted to parameterize the relationship between the subsurface temperature and thermocline depth. See Panel (a) of Figure \ref{Schematic} for a schematic illustration of the model structure and the key components. Note that, the originally pioneering Zebiak and Cane \cite{zebiak1987model} ICM was not designed for characterizing the ENSO complexity. A revised version was recently developed that captures certain diversity features of the ENSO \cite{geng2022enso}. Nevertheless, the new simple ICM to be developed in this paper is very different from the Zebiak and Cane \cite{zebiak1987model} model and its revised version. In fact, the new model highlights the interactions between variabilities in different time scales, where only a minimum nonlinearity is adopted to maintain the model in a simple fashion. The model also exploits suitable stochastic processes to effectively characterize the dynamical properties and accurately reproduce the non-Gaussian statistics of the ENSO complexity in different Ni\~no regions across the equatorial Pacific. The latter is particularly crucial to realistically simulate different ENSO events. It is a key prerequisite for the unbiased statistical forecast of the ENSO complexity as well \cite{majda2018model, fang2022quantifying}.

The rest of the paper is organized as follows. Section \ref{Sec:Model} presents the details of the simple stochastic ICM, including the deterministic and linear interannual components, the stochastic intraseasonal parameterization, the stochastic decadal process, the seasonal synchronization and the nonlinearly coupled multiscale system. Section \ref{Sec:Data} contains the observational datasets and the definitions of different types of ENSO events. The model simulations are presented in Section \ref{Sec:Results} and are compared with the observations. In addition to showing the spatiotemporal patterns of different ENSO events, the skill of reproducing several key statistics in different Ni\~no regions and the composite analysis are also highlighted in this section. Finally, Section \ref{Sec:Conclusion} contains the conclusions and discussion.

\section{The Simple Stochastic ICM}\label{Sec:Model}

\subsection{The starting deterministic and linear interannual model}\label{Sec:Interannual_Model}
The starting interannual model is a deterministic and linear coupled atmosphere-ocean-SST system:

\noindent Atmosphere:
\begin{equation}\label{Starting_Atm}
\begin{split}
&-yv-\partial_{x}\theta=0\\
&yu-\partial_{y}\theta=0\\
&-(\partial_{x}u+\partial_{y}v)=E_{q}/(1-\overline{Q})
\end{split}
\end{equation}
\noindent Ocean:
\begin{equation}\label{Starting_Ocn}
\begin{split}
&\partial_{t}U-c_{1}YV+c_{1}\partial_{x}H=c_{1}{ \tau_{x}}\\
&YU+\partial_{Y}H=0\\
&\partial_{t}H+c_{1}(\partial_{x}U+\partial_{Y}V)=0
\end{split}
\end{equation}
\noindent SST:
\begin{equation}\label{Starting_SST}
\begin{split}
&\partial_{t}T =-c_1\zeta E_{q}+c_1\eta_1 H + c_1\eta_2U.
\end{split}
\end{equation}
The coupled system \eqref{Starting_Atm}--\eqref{Starting_SST} consists of a non-dissipative Matsuno–Gill type atmosphere model \cite{matsuno1966quasi,gill1980some}, a simple shallow-water ocean model \cite{vallis2016geophysical} and a SST budget equation \cite{jin1997equatorial2}.
Here, the state variables $u$ and $v$ are the zonal and meridional wind speeds, $\theta$ is the potential temperature, $U$ and $V$ are the zonal and meridional ocean currents, $H$ is the thermocline depth and $T$ is the SST. All of them are anomalies. For the coordinate variables, $t$ is the interannual time coordinate and $x$ is the zonal coordinate, while $y$ and $Y$ are the meridional coordinates for the atmosphere and ocean components, respectively. The reason to adopt two distinct meridional axes is that the deformation radii of atmosphere and ocean are different.
In these equations, $E_q=\alpha_q T$ is the latent heat with $\overline{Q}$ a constant representing the background vertical moisture gradient \cite{majda2009skeleton}, $\tau_x=\gamma u$ is the wind stress, $\zeta$ is the latent heating exchange coefficient, $\eta_1$ and $\eta_2$ are the strengths of the thermocline and zonal advective feedback, respectively. Here, $\eta_1$ is stronger in the EP due to the shallower thermocline there while $\eta_2$ is stronger in the CP because of the larger zonal gradient of the background SST in that region. The constant $c_1$ is related to the ratio between the ocean and atmosphere phase speeds.
The atmosphere extends over the entire equatorial belt $0\leq x\leq L_A$ with periodic boundary conditions, namely $u(0,y,t)=u(L_A,y,t)$, and similar for other atmospheric variables. The Pacific ocean extends over $0\leq x\leq L_O$ with reflection boundary conditions $\int_{-\infty}^{\infty}U(0,Y,t)\mathrm{d} Y=0$  and $U(L_O,Y,t)=0$ \cite{cane1981response, jin1997equatorial2}.
The detailed parameter values are listed in Appendix.

The above model retains a few essential ingredients that couple the interannual atmosphere, ocean and SST components and drive the basic ENSO dynamics in a simple fashion. Specifically, the latent heating $E_q$ that is proportional to the SST $T$ is depleted from the ocean and forces an atmospheric circulation. The resulting zonal wind stress $\tau_x$ in turn forces ocean dynamics that provides a feedback to the SST through the thermocline depth anomalies $H$ and the zonal current $U$. See the dashed box in Panel (a) of Figure \ref{Schematic} that depicts the interannual components.

To facilitate the computation of the model solution, a meridional projection and truncation is applied to the coupled system, which is known to have meridional bases in the form of parabolic cylinder functions \cite{majda2003introduction, thual2016simple}. For the purpose of developing a simple ICM, only the leading basis function is kept for both the atmosphere  and the ocean, which are denoted by $\phi_0(y)$ and $\psi_0(Y)$, respectively. Both $\phi_0(y)$ and $\psi_0(Y)$ have Gaussian profiles and are centered at the equator but the meridional span of $\phi_0(y)$  is larger than that of $\psi_0(Y)$. The details of these basis functions are included in  Appendix. The meridional truncations trigger atmosphere Kelvin, Rossby waves $K_A, R_A$, and ocean Kelvin, Rossby waves $K_O, R_O$. Once the system \eqref{Starting_Atm}--\eqref{Starting_SST} is projected to the leading meridional bases, the dependence of $y$ and $Y$ are eliminated. The resulting system is only a function of $t$ and $x$. It reads:

\noindent Atmosphere:
\begin{equation}\label{Atmosphere_model}
\begin{split}
&\partial_x K_A  = -\chi_A E_q(2-2\bar{Q})^{-1}\\
  &-\partial_x R_A/3  = -\chi_A E_q(3-3\bar{Q})^{-1}\\
  (B.C.)~~~&K_A(0,t) = K_A(L_A,t)\\
(B.C.)~~~ &R_A(0,t) = R_A(L_A,t)
\end{split}
\end{equation}
\noindent Ocean:
\begin{equation}\label{Ocean_model}
\begin{split}
&\partial_t K_O + c_1\partial_x K_O  = \chi_O c_1\tau_x/2\\
 &\partial_t R_O - (c_1/3)\partial_x R_O  = -\chi_O c_1\tau_x/3\\
 (B.C.)~~~&K_O(0,t) = r_WR_O(0,t)\\
(B.C.)~~~&R_O(L_O,t) = r_EK_O(L_O,t)
\end{split}
\end{equation}
\noindent SST:
\begin{equation}\label{SST_model}
\begin{split}
 &\partial_t T = - c_1\zeta E_q+ c_1\eta_1 (K_O + R_O)+c_1\eta_2 (K_O - R_O),
\end{split}
\end{equation}
where $r_W$ and $r_E$ are the reflection coefficients associated with the ocean reflection boundary conditions (B.C.) while $\chi_A$ and $\chi_O$ are the meridional projection coefficients. Once these waves are solved, the physical variables can be reconstructed,
\begin{equation}\label{physical_reconstruction}
\begin{split}
&u = (K_A - R_A)\phi_0 + (R_A/\sqrt{2})\phi_2\\
&\theta = -(K_A + R_A)\phi_0 - (R_A/\sqrt{2})\phi_2\\
&U = (K_O - R_O)\psi_0 + (R_O/\sqrt{2})\psi_2\\
&H = (K_O + R_O)\psi_0 + (R_O/\sqrt{2})\psi_2
\end{split}
\end{equation}
where $\phi_2$ and $\psi_2$ are the third meridional bases of atmosphere and ocean, respectively.

After applying a spatial discretization in the $x$ direction, the coupled system \eqref{Atmosphere_model}--\eqref{SST_model} is solved numerically via an upwind finite difference scheme.
Since the coupled system is linear and deterministic, its final solution, after the numerical discretization, can be written as a superposition of a set of non-interacting linear modes (the so-called linear solutions). Each linear solution is associated with one eigenmode of the system. In the numerical discretization here, the entire equatorial band is divided into $N_A$ equidistance grids and correspondingly there are $N_O$ grid points in the Pacific ocean. In the simulations of this paper, $N_A=128$ and $N_O=56$ are utilized. In other words, the distance between every two grid points is $312.5$km, as the entire equatorial band and the span of the Pacific ocean are $40,000$km and $17,500$km, respectively.  With an appropriate choice of the physical parameters (see Appendix), all the eigenvalues have negative real parts, indicating the decaying nature of these linear solutions. It is important to highlight that although the eigenvalues of the strongly decaying and fast oscillating small-scale modes may vary by changing the resolution in the spatial discretization, the leading two eigenmodes with the slowest decaying rate remain almost unchanged as long as the spatial discretization is not too coarse. The leading two eigenmodes appear as a pair and the associated eigenvalues are complex conjugate to each other, where the associated oscillation frequency lies on the interannual time scale. Due to the slowest decaying rate, the full solution of the coupled system \eqref{Atmosphere_model}--\eqref{SST_model} is dominated by these eigenmodes.

Figure \ref{Linear_Solutions} shows the spatiotemporal evolution of the leading two eigenmodes. For the purpose of illustration, the decaying rate is manually set to be zero in demonstrating the spatiotemporal pattern of such linear solutions in this figure. By varying the strength of the zonal advective feedback coefficient $\eta_2$, the dominant eigenmodes can have distinct behavior (see Appendix for the detailed parameter values). Specifically, if the role of the zonal advection is weakened, then the leading linear solution exhibits spatiotemporal patterns with the EP El Ni\~no being dominated. See Panel (a). In such a situation, the thermocline feedback is the main mechanism for the SST development. In addition, the convergence center of the atmospheric wind lies in the eastern Pacific. On the other hand, if the zonal advective feedback becomes stronger, then the CP El Ni\~no pattern becomes dominant. Correspondingly, the ocean zonal current leads to the warming in the CP region and the associated convergence center of the atmospheric wind shifts westward as well. See Panel (b). It is worthwhile to highlight that the occurrence frequency of the CP events (every 2.5 years) is higher than that of the EP events (every 4.5 years) in these dominant linear solutions, which is consistent with the observations. Note that these two linear solutions are the necessary conditions and mechanisms for the model to capture the ENSO complexity.

\subsection{Simple stochastic models for the intraseasonal and decadal variabilities}\label{Sec:Intraseasonal_Decadal_Model}
As was seen in Section \ref{Sec:Interannual_Model}, the coupled system \eqref{Atmosphere_model}--\eqref{SST_model} can generate basic linear solutions that exhibit regular patterns of the EP and CP El Ni\~no events in different situations. However, the irregularity and complexity of ENSO require extra mechanisms beyond the deterministic and linear dynamics. In particular, the ENSO variability is often triggered or inhibited by a broad range of random atmospheric disturbances in the tropics, such as the WWBs \cite{harrison1997westerly, vecchi2000tropical, tziperman2007quantifying}, the easterly wind bursts (EWBs) \cite{hu2016exceptionally}, as well as the convective envelope of the MJO \cite{hendon2007seasonal}. On the other hand, it has been shown that the EP and CP events were alternatively prevalent every 10 to 20 years over the past century \cite{yu2013identifying, dieppois2021enso}. For example, the EP events were dominant in the 1980s while the CP El Ni\~nos were more frequently identified after 2000 \cite{chen2015strong, freund2019higher}. These findings imply that the decadal variability plays a crucial role in driving the transitions between the CP- and EP-dominant regimes. Thus, it is also essential to incorporate the decadal effect into the coupled ENSO model to link the separate linear solutions.

To this end, two stochastic processes are developed and coupled to the starting interannual model \eqref{Atmosphere_model}--\eqref{SST_model}. These two stochastic processes characterize the intraseasonal random wind bursts and the decadal variability, respectively. The former is a natural component that depicts  the random atmospheric disturbances. The latter describes the decadal variation of the background Walker circulation. It is related to the climate change scenario and
plays an important role in modulating the strength and the occurrence frequency of the EP and the CP events \cite{chen2015strong,chen2022multiscale}.

First, with the stochastic wind bursts, the wind stress $\tau_x$ now contains two components $\tau_x =\gamma(u+u_p)$, where $u$ remains the same as the atmospheric circulation in \eqref{Starting_Ocn} while $u_p$ is the contribution from the stochastic wind bursts, which is assumed to have the following structure,
\begin{equation*}
  u_{p}(x,y,t)=a_{p}(t)s_{p}(x)\phi_0(y),
\end{equation*}
where $\phi_0(y)$ is again the leading meridional basis while $s_p(x)$ is a fixed spatial structure that is localized in the western Pacific, due to the fact that most of the observed wind bursts are active in the western Pacific. The time series $a_p(t)$ describes the wind burst amplitude and is governed by a simple one-dimensional real-valued stochastic process \cite{gardiner2009stochastic}
\begin{equation}\label{Wind_eqn}
  \frac{\mathrm{d} a_p}{\mathrm{d} t}=-d_{p}a_{p}+{\sigma_{p}(T_{C})}\dot{W}_p,
\end{equation}
where $d_p$ is the damping term that is chosen such that the decorrelation time of the wind is about 1 month. In \eqref{Wind_eqn}, $\dot{W}_p$ is a white noise source while $\sigma_{p}(T_{C})$ is its strength. When $a_p$ is positive and negative, it represents the WWB and EWB, respectively. It is important to highlight that the noise strength ${\sigma_{p}(T_{C})}$ is state-dependent (the so-called multiplicative noise), as a function of the interannual SST from \eqref{SST_model} averaged over the western-central Pacific, namely Ni\~no 4 region. The reason of choosing such a state-dependent noise coefficient is that wind burst activity is usually more active with warmer SST in the western-central Pacific due to the strengthening or eastward extension of the warm pool \cite{vecchi2000tropical, hendon2007seasonal}, which is modeled here in a simple parameterized fashion. Note that the enhanced SST only increases the amplitude of the wind bursts while the individual wind burst event generated from the stochastic process in \eqref{Wind_eqn} does not have a preference towards westerly or easterly. This allows an equal chance to create both the WWB and the EWB as individual events, which is consistent with the observations. Due to the state-dependent noise coefficient, the modeling procedure here indicates that the intraseasonal wind bursts not only affect the interannual variability but also are modulated by the latter as well.

Next, the decadal variability is driven by another simple stochastic process,
\begin{equation}\label{I_eqn}
\frac{\mathrm{d} I}{\mathrm{d} t}  = -\lambda (I - m) + \sigma_I(I) \dot{W}_I,
\end{equation}
where the damping $\lambda$ is set to be $5$ years$^{-1}$ representing the decadal time scale. Similar to \eqref{Wind_eqn}, $\sigma_I(I)$ and $\dot{W}_I$ here are the state-dependent noise strength and the white noise source. The reason to adopt a state-dependent noise coefficient, which is a function of $I$ itself, is to allow the distribution of $I$ to be non-Gaussian. In particular, the trade wind in lower level of the Walker circulation in the decadal time scale is easterly, which means the sign of $I$ should stay the same throughout time and thus the distribution of $I$ is obviously non-Gaussian. This feature can be easily incorporated into the process of $I$ with the state-dependent noise coefficient \cite{averina1988numerical, yang2021enso}. Based on the limited observational data and the theory of inferring the least unbiased maximum entropy solution for a distribution, a uniform distribution between $[0,1]$ is adopted for $I$ in this work. Here, a larger $I$ corresponds to a stronger easterly trade wind. The details of the maximum entropy solution and the way to determine $\sigma_I(I)$ are included in Appendix.
Note that the decadal variability $I$ also stands for the zonal SST gradient between the WP and CP regions that directly determines the strength of the zonal advective feedback, which is the main interaction between decadal and interannual variabilities in the coupled system.
In fact, in Kang et al. \cite{kang2020walker}, a Walker circulation strength index is defined as the sea level pressure difference over the CP/EP region ($160^o$W-$80^o$W, $5^o$S-$5^o$N) and over the Indian ocean/WP region ($80^o$E-$160^o$E, $5^o$S-$5^o$N). The monthly zonal SST gradient between the WP and CP region is highly correlated with this Walker circulation strength index (correlation coefficient of $\sim0.85$), suggesting significant air–sea interaction over the equatorial Pacific. Since the latter is more directly related to the zonal advective feedback strength over the CP region, the decadal variable mainly illustrates such a feature.

\subsection{Seasonal synchronization}
Seasonal phase locking is one of the remarkable features of ENSO, which manifests in the tendency of ENSO to peak during boreal winter and is mainly related to the pronounced seasonal cycle of the mean state \cite{tziperman1997mechanisms,stein2014enso}. The seasonal synchronization is incorporated into the multiscale coupled model developed above through two simple parameterizations.

First, it has been shown that the climatological SST in the central-eastern Pacific cools in boreal fall and warms in spring as a result of the seasonal motion of the Intertropical Convergence Zone (ITCZ), which also modulates the strength of the upwelling and horizontal advection processes to influence the evolution of the SST anomalies \cite{mitchell1992annual}. Since the cool (warm) SSTs tend to coincide with decreased (increased) convective activity and upper cloud cover, a season-dependent damping term, which represents the cloud radiative feedback, is included for describing such a seasonal variation \cite{thual2017seasonal}. More specifically, two sinusoidal functions are utilized for parameterizing the otherwise constant $\alpha_q$, which appears as $E_q=\alpha_q T$ in \eqref{SST_model}. One sinusoidal function has a period of one year that naturally describes the seasonal cycle. The other sinusoidal function has a period of half a year that represents a semiannual contribution to the seasonally modulated variance, as was suggested by Stein et al. \cite{stein2014enso}.

Second, the increased wind burst activity in the western Pacific during the boreal winter as a direct response to the increased atmospheric intraseasonal variability, such as the MJO, is another main contributor of the seasonal synchronization  \cite{hendon2007seasonal, seiki2007westerly}. Therefore, a sinusoidal function with period of one year is utilized for parameterizing the seasonal variation of the wind burst strength coefficient $\sigma_p$ in \eqref{Wind_eqn}.

\subsection{The nonlinearly coupled multiscale system}
The coupled model developed so far is a linear model, despite the state-dependent noise. However, the linear nature of the model is insufficient in characterizing some of the key observed dynamical and statistical features of the ENSO complexity.

From the dynamical point of view, there are at least two major nonlinearities that are expected to be added to the starting linear model. First, the decadal variability  determines the strength of the zonal advective feedback. Therefore, it is natural to treat the modulation of the decadal variability on the ENSO dynamics to be nonlinear, where the decadal variability plays the role of a multiplicative factor of the zonal advection coefficient. In other words, the decadal variability $I$ is incorporated into the SST budget equation \eqref{SST_model} and appear in front of the zonal advection coefficient,
\begin{equation}\label{Final_SST_eqn}
  \partial_t T = - c_1\zeta E_q+ c_1\eta_1 (K_O + R_O)+c_1I\eta_2 (K_O - R_O)+c_1\eta_2c_2,
\end{equation}
such that a quadratic nonlinearity is introduced from $I(K_O - R_O)$ (recall from \eqref{physical_reconstruction} that $K_O$ and $R_O$ are the linear combination of $H$ and $U$). This nonlinearity represents the mechanism that the strengthening of the Walker circulation in the decadal time scale will trigger more CP events. It is crucial in simulating the correct occurrence frequencies of both the CP and the EP El Ni\~nos. One additional small constant $c_2$ is further added to \eqref{Final_SST_eqn}, which guarantees all the variables have climatology with zero mean since otherwise the nonlinearity can cause a slight shift of the mean state.

Another nonlinearity incorporated here is the damping coefficient in the SST equation. Recall that $E_q=\alpha_q T$ and therefore $-c_1\zeta\alpha_q$ is the damping coefficient. Here $\alpha_q$ is parameterized by a quadratic nonlinear function of the CP SST and the spatial structure of such a nonlinear function is concentrated in the CP area. In addition, the symmetric axis of this quadratic function has a negative value, which means a stronger damping is imposed when the CP SST is positive. This effectively gives a cubic damping in the CP region. The reason to introduce this nonlinearity is twofold. On one hand, the relationship between the subsurface temperature and the thermocline depth is more complicated in the CP region \cite{zhao2021breakdown} while only a simple shallow water model is utilized here. Thus, such a nonlinear damping is introduced to parameterize the additional relationship beyond the capability of the shallow water model. On the other hand, it is justified from a simple statistical analysis of the observational data that a linear relationship between the damping and SST anomaly in the CP region is broken while a cubic nonlinearity fits the data in a more accurate fashion \cite{chen2022multiscale}. The nonzero symmetric axis in parameterizing $\alpha_q$ is also crucial for recovering the correct non-Gaussian statistics of the SST in the CP region.

\section{Observational Data Sets and the Definitions of Different Types of the ENSO Events}\label{Sec:Data}
\subsection{Data}
The monthly SST data is taken from the GODAS dataset \cite{behringer2004evaluation}. Anomalies are calculated by removing the monthly mean climatology of the entire period. The Ni\~no 4, Ni\~no 3.4, and Ni\~no 3 indices are the average of SST anomalies over the zonal regions $160^o$E-$150^o$W, $170^o$W-$120^o$W and $150^o$W-$90^o$W, respectively, together with a meridional average over $5^o$S-$5^o$N.

\subsection{Definitions of different types of the ENSO events}\label{Subsec:Def_Events}
The definitions of different El Ni\~no and La Ni\~na events for studying the ENSO complexity follow those in Kug et al. \cite{kug2009two}, which are based on the average SST anomalies during the boreal winter (December–January–February; DJF). When the EP is warmer than the CP and is greater than $0.5^o$C, it is classified as the EP El Ni\~no. Among this, based on the definitions used by  Wang et al. \cite{wang2019historical}, an extreme El Ni\~no event corresponds to the situation that the maximum of EP SST anomaly from April to the next March is larger than $2.5^o$C. When the CP is warmer than the EP and is larger than $0.5^o$C, the event is then defined as a CP El Ni\~no. Finally, when either the CP and EP SST anomaly is cooler than $-0.5^o$C, it is defined as a La Ni\~na event.

\section{Model Simulation Results}\label{Sec:Results}
The numerical solution of the model is calculated utilizing the forward Euler time integration scheme with a time step of $0.5$ days for the interannual variabilities. The Euler-Maruyama scheme is adopted to compute the stochastic processes of the decadal variability and the intraseasonal wind bursts, with a numerical integration time step being $0.5$ days and $0.05$ days, respectively. The monthly averaged model output for the interannual and decadal variabilities is utilized in presenting the dynamical and statistical results. This has almost no difference with the direct model solution but is adopted mainly for the purpose of being consistent with the monthly averaged observational data. On the other hand, the monthly average is not applied to the wind burst data.

\subsection{Model simulation of the ENSO complexity}
\begin{figure}
    \centering\hspace*{-2cm}
    \includegraphics[width=17cm]{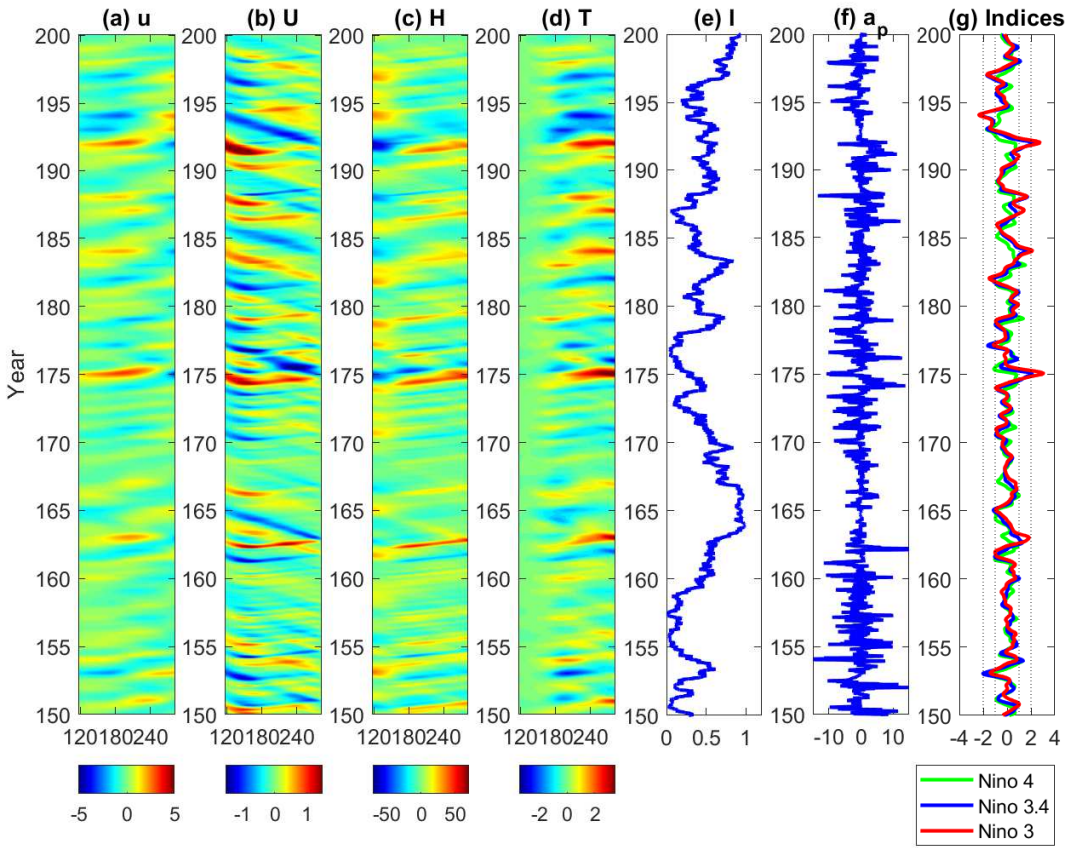}
    \caption{A 50-year model simulation of different fields. Panels (a)--(d): Hovmoller diagrams of the interannual atmosphere wind $u$ (unit: m/s), the ocean current $U$ (unit: m/s), the thermocline depth $H$ (unit: m) and the SST $T$ (unit: $^o$C). The longitude ranges from $120^o$E (120) to $80^o$W (280). Panel (e): time series of the decadal variability $I$. Panel (f): time series of the intraseasonal random wind bursts $a_p$ (unit: m/s). Panel (g): Ni\~no 4, Ni\~no 3.4 and Ni\~no 3 indices. }
    \label{Hovmoller_All_Fields}
\end{figure}

Figure \ref{Hovmoller_All_Fields} shows a 50-year model simulation. With the random and nonlinear components in the model, the resulting atmosphere-ocean-SST fields exhibit irregular spatiotemporal patterns, mimicking the observations shown in Panel (b) of Figure \ref{Schematic}.
To begin with, the model simulation succeeds in reproducing both the realistic CP (e.g., years 154, 157, 160) and EP (e.g., years 163, 175, 192) events as well as some mixed events (e.g., year 188). In fact, the two separate linear solutions presented in Figure \ref{Linear_Solutions} are now linked by the decadal variability, which directly modulates the strength of the zonal advective feedback. In other words, the decadal variability advances the model to have a preference towards either the EP or the CP mode at each time instant, although the details of each single ENSO event are still largely affected by other stochastic and nonlinear effects.
Next, the spatiotemporal fields of the interannual atmosphere wind $u$, ocean current $U$, thermocline depth $H$ and the SST $T$ in Panels (a)--(d) as well as the wind bursts strength $a_p$ in Panel (f) reveal distinct formation mechanisms for the CP and EP El Ni\~nos. The EP El Ni\~no, especially the extreme EP El Ni\~no, is triggered by the random wind bursts and the thermocline depth plays an important role in the event development. In contrast, the zonal advection is the dominant contributor to the CP events. In addition to the response of the CP and EP events to the zonal advective and thermocline feedbacks, the convergence center of the interannual atmosphere wind locates in the CP and EP regions when these two types of the events occur, respectively. These causal relationships are consistent with observations and the previous findings \cite{kao2009contrasting,kug2009two,xiang2013new,zheng2014asymmetry,chen2018observations}. Furthermore, as in the observations, the strength of the CP El Ni\~nos is overall weaker than that of the EP ones \cite{zheng2014asymmetry}. Particularly, extreme El Ni\~no events are only observed in the EP region, due to the anomalously strong wind bursts.
It is also noticed that the probability of generating CP events increases as the decadal variability becomes stronger. This is again consistent with the observations, for example, the CP events becoming more frequent as the strengthening of the Walker circulation in the 21st century \cite{mcphaden2011nino,xiang2013new}. Nevertheless, regardless of the strength of the decadal variability, the model always allows both the CP and the EP events to be triggered with a certain chance. Finally, the Ni\~no indices shown in Panel (f) mimic the reality, where the Ni\~no 4 index has a slightly larger value than Ni\~no 3 at the CP El Ni\~no phases while the Ni\~no 3 index becomes much bigger than Ni\~no 4 during the occurrence of the extreme EP events.

\begin{figure}
    \centering
    \includegraphics[width=12cm]{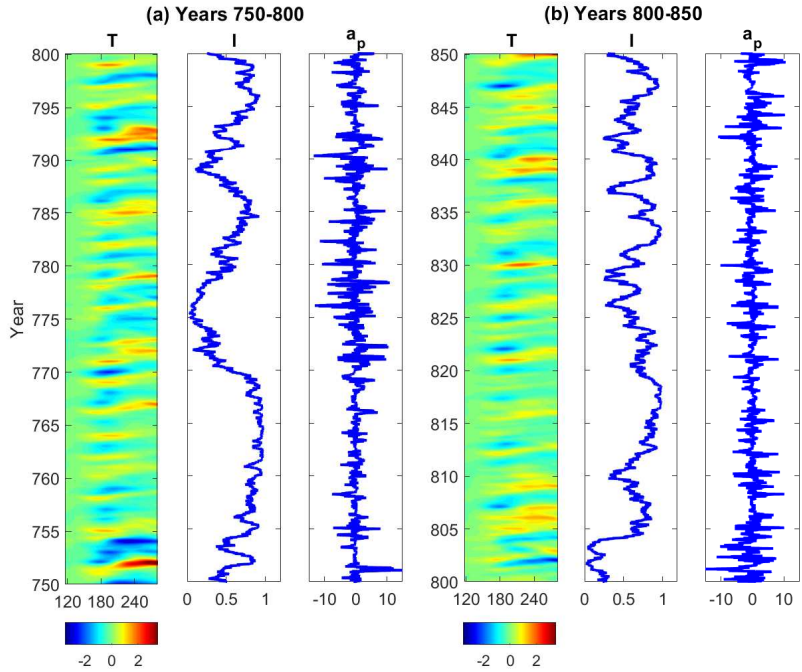}
    \caption{Model simulation: Hovmoller diagram of SST $T$ (unit: $^o$C), time series of the decadal variability $I$ and time series of the intraseasonal random wind bursts $a_p$  (unit: m/s) from year 750 to year 800 and from year 800 to year 850.}
    \label{Hovmoller_SST}
\end{figure}

\begin{figure}
    \centering
    \includegraphics[width=12cm]{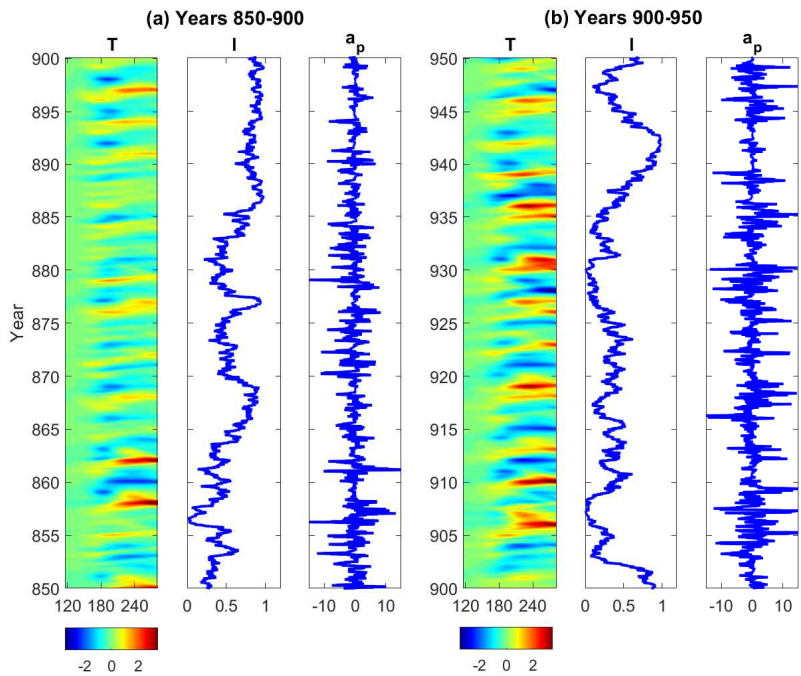}
    \caption{Similar to Figure \ref{Hovmoller_SST} but for the periods from year 850 to year 900 and from year 900 to year 950. }
    \label{Hovmoller_SST2}
\end{figure}

Figures \ref{Hovmoller_SST}--\ref{Hovmoller_SST2} show a simulation of the SST field for 200 consecutive years, accompanying with the associated wind bursts and the decadal variability. To summarize the findings in these figures, Table \ref{Table_Events} lists examples of different ENSO events belonging to 9 refined categories in such a long model simulation and are compared with observations. The results indicate the ability of the model in reproducing the realistic ENSO complexity. First, the model is able to simulate various EP El Ni\~no events with different strengths. In addition to the moderate EP El Ni\~nos, the extreme El Ni\~no events, which appear as a result of the strong WWBs generated from the intraseasonal model, are also reproduced by the model. It is worthwhile to highlight that the so-called delayed super El Ni\~no, as was observed in 2014-2015 \cite{hu2016exceptionally,capotondi2018nature,thual2019statistical,xie2020unusual}, are realistically simulated by the model, for example during model years 905-906. In fact, the model succeeds in recovering the associated peculiar westerly-easterly-westerly wind burst structure that is the key mechanism to trigger such an El Ni\~no event. Here, the initial WWB tends to trigger a strong El Ni\~no but the subsequent EWB kills the event and postpones it until the next year when another series of strong WWBs occur.
Next, the model generates many realistic CP El Ni\~no events. In particular, both single-year (e.g., years 764 and 799) and multi-year (e.g., years 760-761 and 899-900) CP El Ni\~no events can be reproduced from the model. The latter mimics the observed CP episodes, for example during 2018-2020. In addition to those events that clearly belong to either the EP or the CP categories, the model also creates some mixed CP-EP events (e.g., years 785 and 939), which are similar to the observed ones in early 1990s (e.g., the one occurred in year 1992). Finally, the La Ni\~na events from the model usually follow the El Ni\~no ones as the discharge phase. Some La Ni\~na events have cold SST in the CP region while others have cold centers locating around the EP area. The model is also able to simulate multi-year La Ni\~na events, namely, a La Ni\~na transiting to another La Ni\~na, such as the one spans over years 774-775 and 902-904, mimicking the observed events 1999-2000 and 1984-1986, respectively.

\subsection{Comparison of the statistics between model simulations and observations}
In addition to the dynamical properties, the model statistics is another important measurement for assessing its skill in reproducing the realistic ENSO features. Since the focus is on the ENSO complexity, it is essential to study various statistics that represent unique aspects of the ENSO characteristics in different Ni\~no regions across the equatorial Pacific. To this end, four statistical quantities with respect to the SST anomalies of the model simulation are compared with those of the observations in Ni\~no 4, Ni\~no 3.4 and Ni\~no 3 regions, respectively. They are 1) the probability density function (PDF), 2) the seasonal variance, 3) the power spectrum and 4) the autocorrelation function (ACF). Here, the statistics of the observations are computed based on the observed SST between 1951 and 2020, which contains 70 years. On the other hand, a long simulation with in total 3500 years is utilized for computing the model statistics. The total simulation is divided into 50 non-overlapping subperiods, each having the same length as the observation. The statistics is then computed for each of these 50 subperiods, the difference among which reflects the uncertainty in computing these statistics.

\begin{figure}
    \centering
    \includegraphics[width=15cm]{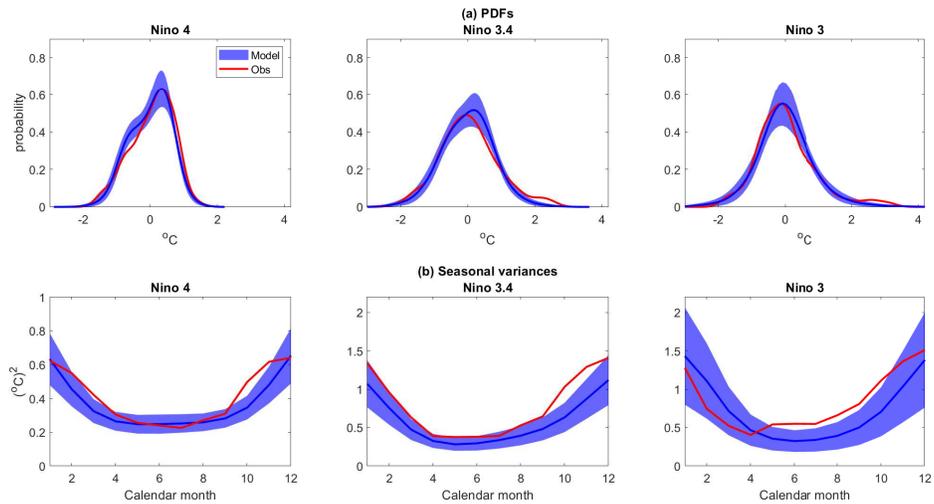}
    \caption{Comparison of the PDFs (Panel (a)) and the seasonal variances (Panel (b)) of SST anomalies between the model simulation and the observations in different Ni\~no regions. The observations are based on the period of 1951-2020, which contains 70 years. Correspondingly, the model simulation has the length in total 3500 years and is divided into 50 equally long periods, each of which is 70 years. The blue shading area is the one standard deviation of the statistics computed from these 50 non-overlapped periods and the blue solid curve is the average value. The observational statistics is shown in red solid curve.}
    \label{Statistics_Results}
\end{figure}

Panel (a) of Figure \ref{Statistics_Results} shows that the model almost perfectly recovers the strong non-Gaussian statistics of the SST anomalies in all the three Ni\~no regions. In particular, the observed Ni\~no 3 SST has a positive skewness and a one-sided fat tail that is associated with the occurrence of the super El Ni\~nos in the EP region. Due to the state-dependent noise in the wind burst process, the model is able to trigger the extreme events and thus accurately reproduce such a highly non-Gaussian distribution. On the other hand, the skewness of the observed Ni\~no 4 SST is negative, and the kurtosis is less than the standard Gaussian value 3, indicating the suppression of extreme El Ni\~no events in the CP region. With the help of the cubic and non-centered damping in the CP area, the model succeeds in capturing such a skewed and light-tailed distribution. Similarly, the model is able to reproduce the PDF of the Ni\~no 3.4 SST, which demonstrates a slight positive skewness. It is worthwhile to highlight that, despite the skill in recovering many dynamical features of ENSO, the CGCMs and many other intermediate models often have difficulties in simultaneously capturing the highly non-Gaussian PDFs in all the three Ni\~no regions, which are nevertheless one of the most important and necessary conditions for simulating the realistic ENSO complexity.
Panel (b) of Figure \ref{Statistics_Results} reveals the capability of the model in recovering the observed seasonal synchronizations of ENSO, which is represented by the monthly variance of the SST in different Ni\~no regions. The observed ENSO events have their onset in boreal spring, develop in summer, and peak in the subsequent winter. These features are overall well captured by the model, especially given the fact that the model only exploits simple sinusoidal functions for parameterizing the seasonal effects.

\begin{figure}
    \centering
    \includegraphics[width=15cm]{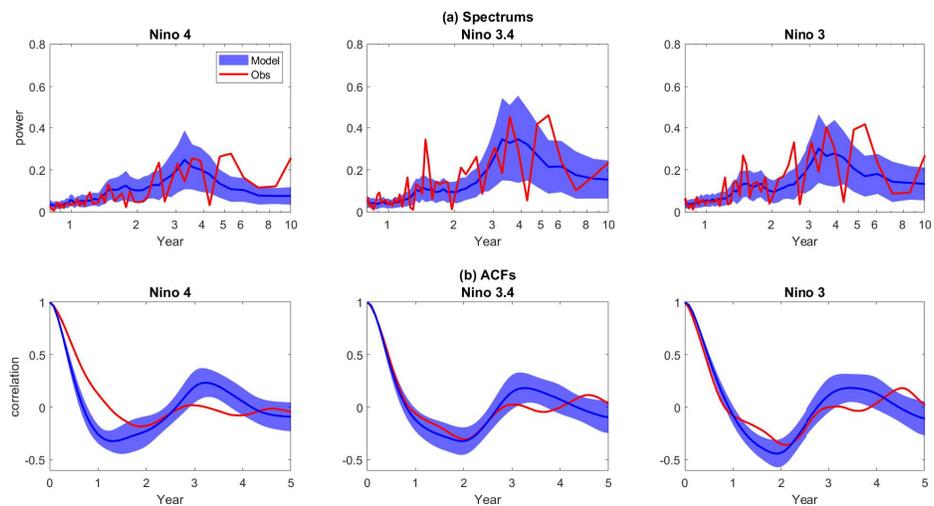}
    \caption{Similar to Figure \ref{Statistics_Results} but for the spectrums (Panel (a)) and the autocorrelation functions (ACFs; Panel (b)).}
    \label{Statistics_Results2}
\end{figure}

Next, Panel (a) of Figure \ref{Statistics_Results2} shows the power spectrums of the SST. It can be seen that the major signal of the power associated with the Ni\~no 4 SST is between 2 and 4 years. The power decreases rapidly outside this window. In contrast, the signal of the Ni\~no 3 (and Ni\~no 3.4) SST has a wider
range, that is, the power remains significant between the range of 2 and 7 years. These features are well captured by the model simulations, which are the key requirements for the model to reproduce the  observed irregular oscillations. In addition, as is shown in Panel (b), the model is able to create very similar ACFs in different Ni\~no regions as the observations. It indicates that the model has realistic decaying rate and memory, which are consistent with nature in the equatorial Pacific and are important prerequisites for the forecast of the ENSO complexity.

\begin{figure}
    \centering
    \includegraphics[width=12cm]{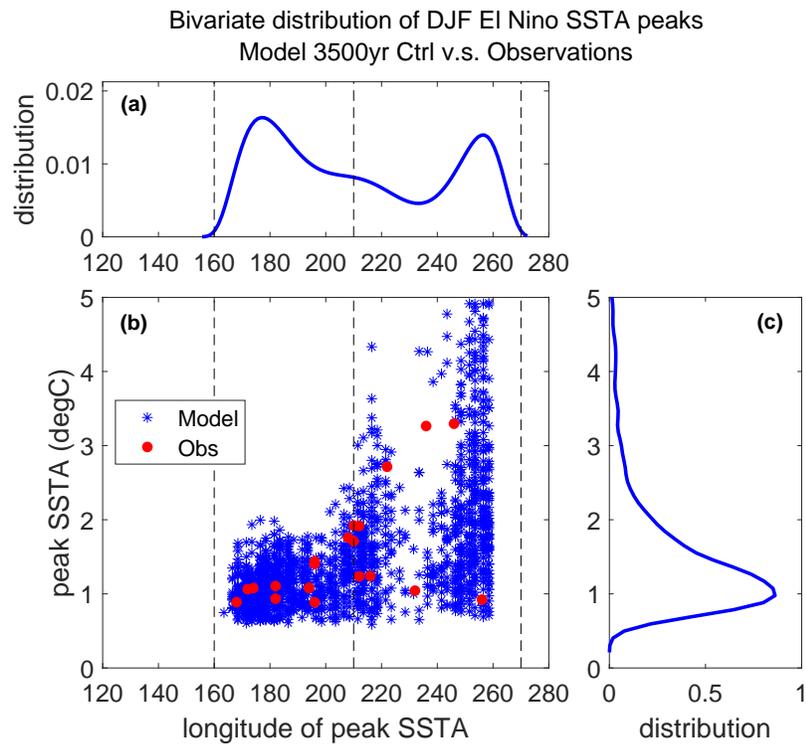}
    \caption{Distribution of equatorial Pacific SST maxima for the El Ni\~no events from the model simulation of 3500 years (blue) and the observations (red). For each of the qualified El Ni\~no events, the winter-mean SST anomalies are averaged over the equatorial zone (from 5$^o$S to 5$^o$N), and then the Pacific zonal maximum is located. (a) Distribution of peak SST anomaly longitudes. (b) Scatter plot of the peak SST anomaly value v.s. the longitude at which it occurs. The blue (red) dots are for the model results (observations). (c) Distribution of peak SST anomaly values.}
    \label{Bivariate_Distribution}
\end{figure}

Finally, Figure \ref{Bivariate_Distribution} shows the scatter plot of the equatorial Pacific boreal winter (DJF) mean SST maxima along the equator for the El Ni\~no events from the model simulation, which is compared with the observations. Each point displayed here is a function of the maximum SST and its corresponding longitude.
Panel (a) shows the distribution of the locations of these El Ni\~no events. The distribution exhibits two major centers. One is near the dateline and another is in the cold tongue region. These two peaks correspond to the CP and EP El Ni\~no events, respectively. It should be noticed that the model distribution here is in general more consistent with the observations than most of the state-of-art CGCMs, which often exhibit a common bias for a farther west position \cite{capotondi2015understanding}. Another desirable feature to highlight is that, despite the bimodality, there remains a relatively large probabilities of the event occurrence in the region spanning from the dateline to $120^o$W (240). This reveals that the ENSO diversity is not simply composed by the events that belong to two separate categories. Instead, there are many mixed EP-CP events. In fact, according to Panel (b), there are several observed El Ni\~no events (red dots) locating in this region, indicating that the distribution should be in the form of a continuum rather than two disjoint sets \cite{johnson2013many}. This, however, seems not the case in many CGCM results, as was pointed out by  Capontondi et al. \cite{capotondi2015understanding}.
Next, in terms of the strength, the events with CP SST anomaly peaks are overall weaker than the corresponding EP ones. While the strongest events always locate in the eastern Pacific, the EP events can exhibit a wide range of amplitudes. These characteristics are physically reasonable since there is more potential for the warming amplitude in the EP due to its climatological SST being much less than the radiative-convective equilibrium temperature of about 30$^o$C while the potential for the warming amplitude in the CP is relatively smaller \cite{jin2003strong}.

\subsection{Composite analysis}
To provide a more quantitative assessment of the model performance on simulating each type of the ENSO events, Table \ref{Table:Frequency} summarizes the occurrence frequency of different El Ni\~no and La Ni\~na events (as were defined in Section \ref{Subsec:Def_Events}) per 70 years. For the El Ni\~no events, although the occurrence frequency of both the EP and the CP El Ni\~no events from the model (18.2 and 14.7) is higher than that from the observations (14 and 10), the gap in counting both types of the El Ni\~nos between the model and the observations is just around one standard deviation of different model simulation segments, which is nevertheless within a relatively reasonable range. Such a difference mainly comes from  overestimating the number of the multi-year El Ni\~no events in the model. Except for this overestimation issue, other statistics from the model simulation are all very similar to those from the observations. First, the ratio between the numbers of the EP and the CP El Ni\~no events from the model simulation (55\% v.s. 45\%) is almost the same as that in the observations (58\% v.s. 42\%), which indicates the skill of the model in capturing the overall El Ni\~no diversity. Second,  a total of four extreme El Ni\~no events have occurred since 1951, namely 1972–1973, 1982–1983, 1997–1998 and 2015–2016, while a comparable number of 3.1$\pm$2.3 events is found in the model simulations. Third, the occurrence frequency for the La Ni\~na events (26.1$\pm$4.6) from the model is very close to that in the observations (24). In particular, the model and the observations share approximately the same numbers of both the single-year and multi-year La Ni\~na events. It is worthwhile to remark that, as the classification of El Ni\~no events in observations is subject to the limited sample size and suffers from uncertainties associated with varying datasets \cite{wiedermann2016climate,capotondi2020enso}, the perfect agreement of the occurrence frequency with observations should not be a strict metric on evaluating the model performance. Therefore, it can be concluded that the model is overall skillful in reproducing reasonably accurate numbers of different types of the ENSO events.

\begin{figure}
    \centering
    \includegraphics[width=10cm]{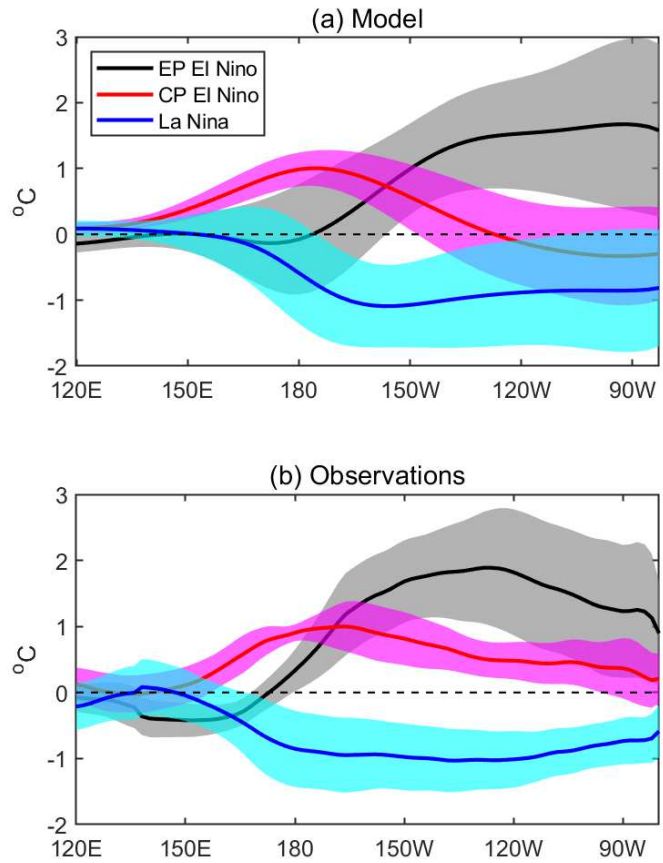}
    \caption{Composited winter (DJF) mean SST anomalies (lines) and the corresponding error bars (i.e., shaded by one standard deviation) over the equatorial Pacific for EP El Ni\~no (black), CP El Ni\~no (red) and La Ni\~na (blue) events. (a) and (b) are for model and observation, respectively.}
    \label{ENSO_composite}
\end{figure}

Next, Figure \ref{ENSO_composite} exhibits the composites of the DJF mean SST anomalies on the equatorial Pacific for the EP El Ni\~no, the CP El Ni\~no and the La Ni\~na with respect to the spatial distribution. Here, all the ENSO events (i.e., during the total 3500 model years) are used to compare with the observations, instead of separating it into the smaller segments. It is seen that the composites from the model simulation are almost identical with those from the observations in terms of both the spatial patterns and the amplitudes. Specifically, the warming center locates in the EP and CP regions for the EP and CP El Ni\~nos, respectively, although the simulated EP events are closer to the coastline of the South America. The cooling center of La Ni\~na is located between the warming centers of the EP and CP El Ni\~no, which is also in accordance with the observations. Next, the model succeeds in recovering the ENSO asymmetry. In other words, the amplitude of the EP El Ni\~no is overall stronger than those of the CP El Ni\~no and the La Ni\~na \cite{hayashi2020dynamics}. It should be noticed that accurately reproducing the spatial distributions of the ENSO events and the ENSO asymmetry is still one of the main challenges for the state-of-art CGCMS \cite{planton2021evaluating}. 

\subsection{Sensitivity analysis}

What remains is to study the role of each key process in the coupled model, which facilitates the understanding of the model dynamics. To this end, several sensitivity tests are carried out in the following.

Let us begin with investigating the role of the decadal variability. In the standard run, $I$ is driven by a simple stochastic process \eqref{I_eqn} and its value of the decadal variability time series varies between $0$ and $1$. In the sensitivity tests, the model simulations with a fixed $I$ of either $I\equiv0$ or $I\equiv1$ are studied. In each test, the strength of the wind bursts is slightly tuned by multiplying a constant such that the variance of the SST remains the same as the standard run, which allows a fair comparison for the occurrence frequency in different scenarios. Note that the decadal variability is related to the climate change and climate projection. Therefore, the response of the interannual variability due to the variation of the decadal variability is of great interest.

First, the decadal variability is set to be zero (i.e., $I\equiv0$), which corresponds to the situation with a weakened background Walker circulation as the period between 1980 and 2000 but towards the more extreme case. In such a scenario, the model simulation leads to an increase of the El Ni\~no events (from 32.9 to 42.1 per 70 years) and a decrease of the La Ni\~na ones (from 26.1 to 19.4). More specifically, among the El Ni\~no events, 65.8\% events are the EP type while only 34.2\% events remain as the CP El Ni\~no. This means the scenario with an weakened Walker circulation is more favorable for the EP than the CP El Ni\~no events, as the zonal advective feedback in this situation is reduced and the role played by the thermocline feedback becomes dominant. Note that the CP El Ni\~no events can still be generated because of the stochastic noise. Similar to the overall occurrence frequency, more multi-year El Ni\~no (from 10.5 to 15.9) and less multi-year La Ni\~na (from 6.9 to 4.6) events are found in such a case. In addition, about 7.6 extreme El Ni\~no events are produced per 70 years. This is nearly twice as many as that in the standard run and is in accordance with the observations, where 3 out of the total 4 extreme El Ni\~no events occurred before 2000. Notably, such a result is consistent with the climate projection that an increased frequency of extreme El Ni\~no events will appear due to the greenhouse warming, since a projected surface warming over the EP is faster than that in the surrounding ocean waters \cite{cai2014increasing}. On the other hand, if the decadal variability is set to be one (i.e., $I\equiv1$), then the model mimics the situation when the Walker circulation and zonal thermocline slope are relatively strong, similar to the period after 2000. In such a scenario, more CP El Ni\~no events (from 10 to 20.6) are found in the model simulation, as a natural consequence of the strengthened zonal advective feedback. Similarly, less multi-year El Ni\~no (from 10.5 to 9.6) and multi-year La Ni\~na (from 6.9 to 5.2) events are generated. In addition, there remains only 1.3 extreme El Ni\~no events per 70 years since the overall occurrence of the EP events becomes lower. These findings further indicate that the difference between the positive and negative phases of ENSO is weakened. As a result, the PDF of the EP SST becomes more towards nearly Gaussian, which is fundamentally different from the observed non-Gaussian PDF with a fat tail.

The next analysis study is about the effect of the multiplicative noise $\sigma_p(T_C)$ in the stochastic wind burst process. The multiplicative noise is one of the main contributors to the asymmetry of the EP type of El Ni\~no. If an additive noise (i.e., setting $\sigma_p$ as a constant) is adopted instead, then the PDF of the simulated EP SST becomes more symmetric and Gaussian. This is very different from the observations, where the amplitude of the extreme El Ni\~no is typically larger than that of the strongest La Ni\~na.

Finally, the nonlinear damping in the CP region is crucial to the ENSO dynamics and statistics. According to the observations, the asymmetry with respect to the SST PDF in the CP is reversed compared with that in the EP. That is, the amplitude of the negative phase of the CP SST is in general stronger than that of the positive one, which leads to a negative skewness of the CP SST PDF. Such a negative skewness is accurately recovered with the help of the nonlinear function of $\alpha_q$ with the non-zero symmetric axis. On the other hand, the nonlinear damping in the CP region also plays an important role in suppressing the amplitude of the strong CP events, since the damping becomes larger as the amplitude of the SST anomaly. As a result, the system is more in favor of the small and moderate SST anomalies and a reduced kurtosis appears for the CP SST PDF.

\section{Conclusions and Discussion}\label{Sec:Conclusion}

In this paper, a simple multiscale stochastic ICM is developed to capture the ENSO diversity and complexity. The model highlights the interconnections between intraseasonal, interannual, and decadal variabilities. It also exploits suitable stochastic processes to facilitate the realistic simulation of the ENSO. The model succeeds in reproducing the spatiotemporal dynamical evolution of different types of the ENSO events. It also accurately recovers the strongly non-Gaussian probability density function, the seasonal phase locking, the power spectrum and the temporal autocorrelation function of the SST anomalies in all the three Ni\~no regions (3, 3.4 and 4) across the equatorial Pacific. Furthermore, both the composites of the SST anomalies for various ENSO events and the strength-location bivariate distribution of equatorial Pacific SST maxima for the El Ni\~no events from the model simulation highly resemble those from the observations. These desirable features of the model are particularly important for the purpose of realistically simulating different ENSO events. They are also the key prerequisites for the unbiased statistical forecast of the ENSO complexity.

It is worthwhile to point out that the stochastic ICM developed here share many common features as the conceptual model in Chen et al. \cite{chen2022multiscale}. Both models include three time scales, where the decadal variability modulates the solution that alternates between the EP- and the CP-dominant regimes while the intrasesonal wind bursts triggers most of the irregularities and extreme events. The underlying principles of incorporating stochastic wind bursts and the nonlinearity into both models also appear in a similar fashion. Nevertheless, the ICM emphasizes more sophisticated physics and includes many additional dynamical properties. It also involves spatially-extended structures, which allow a better understanding and potentially an improved forecast of the spatiotemporal patterns.

A few important topics are remained as future work. First, the intraseasonal model adopted here is a one-dimensional simple stochastic process, where the spatial structure is prescribed and fixed. A more realistic intraseasonal model can be a spatially-extended (stochastic) model for the wind bursts and the MJO, for example one of the models in  Adames et al. \cite{adames2016mjo}, Wang et al. \cite{wang2016trio} and Thual et al. \cite{thual2018tropical}. A dynamical intraseasonal model allows the wind bursts and the MJO to have realistic spatial propagation mechanisms, spanning from the Indian ocean to the WP. Such a coupled model will also be helpful in understanding of the coupling between the MJO and ENSO. Second, the model developed here has a symmetric meridional structure, due to the fact that only the leading meridional basis function is utilized in the meridionally truncated system. Yet, both the wind bursts and ENSO have certain meridionally asymmetric features. Therefore, incorporating additional meridional basis functions into the model is a natural extension of the current system. Third, the ICM developed here is applicable to the forecast of different ENSO events. In particular, the ICM can be combined with the conceptual model in Chen et al. \cite{chen2022multiscale} (and possibly the coupled MJO-ENSO model to be developed as well) for the multi-model data assimilation and forecast, which advances the understanding of the role of each model and each component in improving the forecast of the ENSO complexity.

\bibliography{references}



\section*{Acknowledgements}
The research of N.C. is partially funded by the Office of VCRGE at UW-Madison. The research of X.F. is supported by the Ministry of Science and Technology of the People's Republic of China (Grant No. 2020YFA0608802) and the National Natural Science Foundation of China (Grant No. 41805045).

\clearpage

\section*{Appendix-A: Variables and Parameters}
The definitions and units of the model variables are listed in Table \ref{Table:variables}. The parameter values are summarized in Table \ref{Table:param}.

\begin{figure}
    \centering
    \includegraphics[width=15cm]{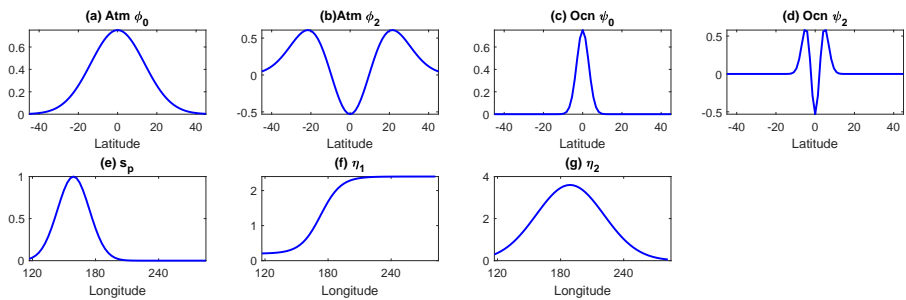}
    \caption{Panels (a)--(d): Spatial structure functions of the meridional bases $\phi_0(y)$, $\phi_2(y)$, $\psi_0(Y)$ and $\psi_2(Y)$. Panels (e)--(g): Spatial structure functions of $s_p(x)$, $\eta_1(x)$ and $\eta_2(x)$.}
    \label{fig:my_label}
\end{figure}

As was stated in Section \ref{Sec:Interannual_Model}, different parabolic cylinder functions in the ocean and atmosphere were used in the coupled model. Their profiles are shown in Figure \ref{fig:my_label}. The atmospheric parabolic cylinder functions read $\phi_0(y)=(\pi)^{-1/4}\exp(-y^2/2)$, $\phi_2(y)=(4\pi)^{-1/4}(2y^2-1)\exp(-y^2/2)$. The ocean parabolic cylinder functions read $\psi_m(Y)$, which have the same profiles as the atmospheric ones but depend on the oceanic meridional axis $Y$.

In order to couple the ocean and atmosphere, projection coefficients are introduced, which read $\chi_A=\int_{ -\infty }^{+\infty} {\phi_0(y)\phi_0(y/\sqrt{c})\mathrm{d} y}$ and $\chi_O=\int_{ -\infty }^{+\infty} {\psi_0(Y)\psi_0(Y/\sqrt{c}Y)\mathrm{d} Y}$. The atmosphere uses a truncation of the Kelvin and first Rossby atmospheric equatorial waves of amplitude $K_A$ and $R_A$. The ocean uses a truncation of zonal wind stress forcing to $\psi_0$, $\tau_x=\tau_x\psi_0$. This is known to excite only the Kelvin and first Rossby oceanic waves, of amplitude $K_O$ and $R_O$. Similarly, for the SST model  a truncation $\psi_0$, $T=T\psi_0$ is utilized. Then, the deterministic and linear part of the ENSO model truncated meridionally yields \eqref{Atmosphere_model}--\eqref{SST_model}.

\section*{Appendix-B: Parameters for the Two Different Linear Solutions}
The model starts with a deterministic and linear coupled interannual atmosphere, ocean and SST system. Before the two stochastic processes on the other two time scales are further incorporated, it is crucial to confirm that the linear model is able to generate the basic solutions of the two types of ENSO events under different parameter settings. In addition, the potential difference should be as small as possible for the physical interpretation.

\begin{figure}[!h]
    \centering
    \includegraphics[width=12cm]{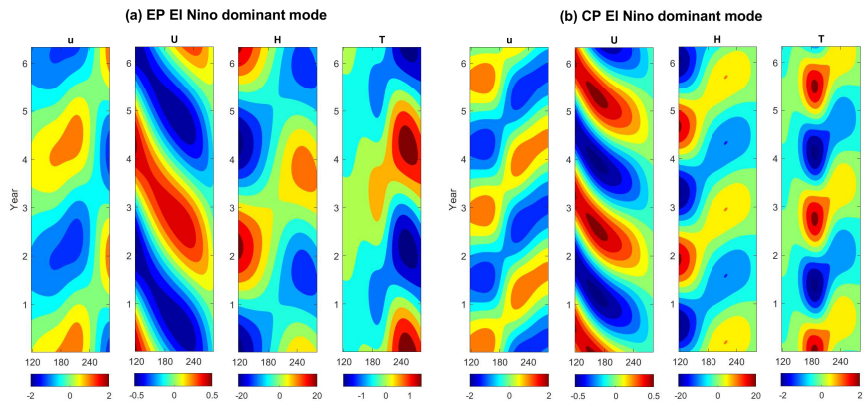}
    \caption{Linear solutions of the coupled system \eqref{Atmosphere_model}--\eqref{SST_model} reconstructed utilizing the leading two eigenmodes, which have the slowest decaying rate. These two modes appear as a pair of complex conjugate and therefore the reconstructed spatiotemporal pattern is real-valued. Panel (a) shows the solution by multiplying a small number to the ocean zonal advective feedback $\eta_2$ coefficient to lower its role and thus gives a EP El Ni\~no dominant mode. Panel (b) shows the solution of the system with a stronger zonal advective feedback and leads to a CP El Ni\~no dominant mode. In both panels, the four columns present the hovmoller diagrams of the interannual atmosphere wind $u$ (unit: m/s), the ocean current $U$ (unit: m/s), the thermocline depth $H$ (unit: m) and the SST $T$ (unit: $^o$C). The detailed parameter values corresponding to the results here are listed in Appendix. Note that, for the purpose of illustration, the decaying rate is manually set to be zero in demonstrating the spatiotemporal pattern of such linear solutions here.}
    \label{Linear_Solutions}
\end{figure}

In this model, the behavior of the leading mode is mainly determined by the relative amplitude between the zonal advective feedback and the thermocline feedback, which is consistent with the observational analyses. For this purpose, the only change of the setting is the strength of the zonal advective feedback, i.e., $\eta_2$. Specifically, for the linear solution corresponding to the CP ENSO regime, $\eta_2(x)=\max(0,4\exp(-(x-L_O/(7/3))^2/0.05)\times0.9)$ is utilized (Panel (b) of Figure \ref{Linear_Solutions}), while for that corresponding to the EP ENSO regime, $\eta_2(x)=\max(0,4\exp(-(x-L_O/(7/3))^2/0.05)\times0.9)\times0.3$ is adopted (Panel (a) of Figure \ref{Linear_Solutions}), which is $30\%$ of the one in the CP ENSO regime.

\section*{Appendix-C: Stochastic Process with Multiplicative Noise for the Decadal Variability}
The decadal variability influences the occurrence frequency of the two types of El Ni\~no and thus the ENSO complexity. A stochastic model is introduced for the decadal variability, which depicts the strength of the background Walker circulation and affects the related zonal advective feedback. In the decadal model \eqref{I_eqn}, a state-dependent (i.e., multiplicative) noise coefficient $\sigma_I(I)$ is adopted that allows  $I$ to be non-negative, which comes from the fact that the trade wind in the lower level of the Walker circulation in the decadal time scale is  easterly. Here, as only limited data for the decadal variability is available, a uniform distribution function of $I$ is adopted in the model. This is based on the fact that the uniform distribution is the maximum entropy solution for a function in the finite interval without additional information \cite{kapur1992entropy, majda2006nonlinear, branicki2013non}. The parameter $m$ is the mean of $I$, which can be inferred directly from the data. The damping parameter $\lambda$ can be determined by taking the inverse of the decorrelation time, which is defined as
\begin{equation*}
  \tau = \lim_{T\to\infty}\int_0^T ACF(t)\mathrm{d} t \qquad\mbox{with}\qquad ACF(t) = \lim_{T\to\infty}\frac{1}{T}\int_0^T\frac{I(t+t')I(t')}{var(I)}\mathrm{d} t'.
\end{equation*}
In practice, a sufficiently large $T$ is taken as a numerical approximation. Finally, the multiplicative noise coefficient $\sigma_I(I)$ is determined in the following way \cite{averina1988numerical}
\begin{equation*}
  \sigma^2_I(I) = \frac{-2\lambda}{p(I)}\int_{-\infty}^I\left(s-\frac{m}{\lambda}\right)p(s)\mathrm{d} s.
\end{equation*}



%
\begin{table}
\centering
\begin{tabular}{llll}
  \hline
  Coarse category & Refined category & Model (yrs 750-950) & Observations (yrs 1980-2020)\\\hline
  EP El Ni\~no& & & \\
  &Moderate EP El Ni\~no & 792, 897 & 1987 \\
  &Super El Ni\~no & 752, 862 & 1998 \\
  &Delayed super El Ni\~no & 918-919, 905-906 & 2014-2015 \\\hline
  CP El Ni\~no& & & \\
  &Single-year CP El Ni\~no & 764, 799 & 2005 \\
  &Multi-year CP El Ni\~no: & 760-761, 899-900 & 2018-2020 \\
  & CP La Ni\~na &770, 848 & 1989\\\hline
  Mixed events & & &\\
  &Mixed EP-CP El Ni\~no & 785, 939 & 1992 \\\hline
  La Ni\~na & & &\\
  &Single-year La Ni\~na & 798, 802 & 2006 \\
  &Multi-year La Ni\~na & 774-775, 902-904 & 1999-2000 \\
  \hline
\end{tabular}\caption{Examples of different ENSO events in the model simulation (from year 750 to year 950; showing in Figures \ref{Hovmoller_SST}--\ref{Hovmoller_SST2}) and observations (from year 1980 to 2020). Here, two examples from the model simulation and one example from observations are listed for each type of the ENSO events, respectively.  }\label{Table_Events}
\end{table}

\begin{table}
 \centering
 \begin{tabular}{l lllllll}
 \hline
  &\multicolumn{4}{c}{El Ni\~no}  & & \multicolumn{2}{c}{La Ni\~na}  \\
 \hline
   & EP & CP & Extreme &Multi-year &\qquad&Total &Multi-year\\
   Obs &  14  &  10 &    4&     5&&    24 &    8\\
   Model  & $18.2\pm3.5$ &  $14.7\pm3.5$  &  $3.1\pm2.3$ &  $10.5\pm1.6$  && $26.1\pm4.6$  &  $6.9\pm2.6$  \\
 \hline
 \end{tabular} \caption{Occurrence frequency of different ENSO events per 70 years. The observations are based on the period of 1951-2020, which contains 70 years. For the model simulation, the mean value plus and minus one standard deviation computed from the 50 segments is shown for each case. Note that the counted number of the EP and the CP El Ni\~nos contains both single-year and multi-year events as well as the extreme events. Therefore, the total number of the El Ni\~no events is simply the summation of the numbers in the first two columns (e.g., 24 in observations).  }\label{Table:Frequency}
 \end{table}
 
\begin{table}
 \centering
 \begin{tabular}{llll}
 \hline
   Variable &  Unit  &  Unit Value\\
   $x$ zonal axis & $[y]/\delta$ & 15000km \\
   $y$ atmospheric meridional axis  & $\sqrt{c_A/\beta}$ & 1500km \\
   $Y$ oceanic meridional axis & $\sqrt{c_O/\beta}$ & 330km \\
   $t$ interannual time axis & $[t]$ & 34days \\
   $u$ zonal wind speed & $\delta c_A$ & 5ms$^{-1}$ \\
   $v$ meridional wind speed & $\delta [u]$ & 0.5ms$^{-1}$ \\
   $\theta$ potential temperature & 15$\delta$ & 1.5K \\
   $E_q$ latent heating & $[\theta]/[t]$ & 0.45K.day$^{-1}$ \\
   $U$ zonal current speed & $c_O\delta_O$ & 0.25ms$^{-1}$ \\
   $V$ meridional current speed & $\delta \sqrt{c}[U]$ & 0.56cms$^{-1}$ \\
   $H$ thermocline depth & $H_O\delta_O$ & 20.8m \\
   $T$ sea surface temperature & $[\theta]$ & 1.5K \\
   $a_p$ wind burst amplitude & $[u]$ & 5ms$^{-1}$ \\
 \hline
 \end{tabular} \caption{Model variables, definitions and units.}\label{Table:variables}
 \end{table}

\begin{table}
 \centering
 \renewcommand{\arraystretch}{0.9}
 \begin{tabular}{lll}
 \hline
   Parameter & Value\\
   $\epsilon$ Froude number & 0.4 \\
   $\delta$ long-wave scaling & 0.1 \\
   $\delta_O$ arbitrary constant & 0.1 \\
   $c_A$ atmospheric phase speed & 50ms$^-1$ \\
   $c_O$ oceanic phase speed & 2.5ms$^-1$ \\
   $c$ ratio of oceanic/atmospheric phase speed & 0.05 \\
   $c_1$ modified ratio of phase speed & 0.15 \\
   $\beta$ beta-plane parameter & 2.28 $10^{-11}$m$^{-1}$s$^{-1}$ \\
   $g'$ reduced gravity & 0.03ms$^{-2}$ \\
   $H_O$ mean thermocline depth & 50m \\
   $\rho_O$ ocean density & 1000 kg.m$^{-3}$ \\
   $\chi_A$ atmospheric meridional projection coefficient & 0.31 \\
   $\chi_O$ oceanic meridional projection coefficient & 1.38 \\
   $L_A$ equatorial belt length & 8/3 \\
   $L_O$ equatorial Pacific length & 1.2 \\
   $\bar{Q}$ mean vertical moisture gradient & 0.9 \\
   $\bar{T}$ mean SST & 16.6 (which is 25$^o$C)\\
   $\alpha_q$ latent heating factor & $\alpha_q=q_cq_e\exp(q_e\bar{T})/\tau_q\times \beta_1(T)\times \beta_2(t)$ \\
   $\beta_1(T)$ state dependent component in $\alpha_q$ & $\beta_1(T) = 1.8- \eta_2/3+(0.2 + |T_C+0.4|\times \eta_2 )^2/5$\\
   $\beta_2(T)$ seasonal dependent component in $\alpha_q$ & $\beta_2(T) = 1+0.5\sin(2\pi(t-1/12)) + 0.1\sin(2\pi t) \eta_2$ \\
   &$\qquad\qquad- 0.0625\sin(4\pi(t-3/12)) \eta_1$\\
   $q_c$ latent heating multiplier coefficient & 7 \\
   $q_e$ latent heating exponential coefficient & 0.093 \\
   $\tau_q$ latent heating adjustment rate & 15 \\
   $\gamma$ wind stress coefficient & 6.53 \\
   $r_W$ western boundary reflection coefficient & 0.5 \\
   $r_E$ eastern boundary reflection coefficient & 1 \\
   $\zeta$ latent heating exchange coefficient & 8.7 \\
   $c_2$ mean correction coefficient & 0.1\\
   $\eta$ profile of thermocline feedback & $\eta(x)=1.3+(1.1\times\tanh(7.5(x-L_O/3)))$ \\
   $\eta_2$ profile of zonal advective feedback & $\eta_2(x)=\max(0,4\exp(-(x-L_O/(7/3))^2/0.1)\times0.9)$ \\
   $d_p$ wind burst damping & 1.12 (which is 1mon$^{-1}$) \\
   $s_p$ wind burst zonal structure & $s_p(x)=\exp(-45(x-L_O/4)^2)$ \\
   $\sigma_p(T_C)$ wind burst noise coefficient & $\sigma_p(T_C)=1.6(\tanh(T_C)+1) (1+0.6\cos(2\pi t))$ \\
   &$(1-0.75I)$ \\
   $\lambda$ damping of decadal variability & 0.0186 (which is 5year$^{-1}$) \\
   $m$ mean of $I$ & 0.5 \\
\hline
\end{tabular} \caption{Model parameter values.}\label{Table:param}
\end{table}

%


\end{document}